\documentclass[conference]{IEEEtran}
\IEEEoverridecommandlockouts
\usepackage{cite}
\usepackage{amsmath,amssymb,amsfonts}
\usepackage{algorithmic}
\usepackage{graphicx}
\usepackage{booktabs,siunitx}
\usepackage{textcomp}
\usepackage{multirow}
\usepackage{multicol}
\usepackage{xcolor}
\usepackage{subcaption}
\usepackage{comment}
\usepackage{enumitem}

\usepackage{tikz}
\usetikzlibrary{bayesnet}
\usetikzlibrary{arrows, automata}
\usetikzlibrary{arrows.meta}
\usetikzlibrary{shapes, arrows, positioning, calc}
\usetikzlibrary{shadows.blur}
\usetikzlibrary{shapes.symbols}

\def\BibTeX{{\rm B\kern-.05em{\sc i\kern-.025em b}\kern-.08em
    T\kern-.1667em\lower.7ex\hbox{E}\kern-.125emX}}

\begin{document}
\bstctlcite{IEEEexample:BSTcontrol}

\title{Joint Training of Speaker Embedding Extractor, Speech and Overlap Detection for Diarization
\thanks{Computing on IT4I supercomputer was supported by the Czech Ministry of Education, Youth and Sports through the e-INFRA CZ (ID:90254).}}

\author{
    \IEEEauthorblockN{Petr P\'{a}lka$^{1}$, Federico Landini$^{1}$, Dominik Klement$^{1}$, Mireia Diez$^{1}$, Anna Silnova$^{1}$, Marc Delcroix$^2$, Luk\'{a}\v{s} Burget$^{1}$}
    \IEEEauthorblockA{
  $^1$Brno University of Technology, Speech@FIT, Czechia\hspace{1cm} $^2$NTT Corporation, Japan
    }
}

\maketitle

\begin{abstract}
In spite of the popularity of end-to-end diarization systems nowadays, modular systems comprised of voice activity detection (VAD), speaker embedding extraction plus clustering, and overlapped speech detection (OSD) plus handling still attain competitive performance in many conditions. However, one of the main drawbacks of modular systems is the need to run (and train) different modules independently. In this work, we propose an approach to jointly train a model to produce speaker embeddings, VAD and OSD simultaneously and reach competitive performance at a fraction of the inference time of a standard approach. Furthermore, the joint inference leads to a simplified overall pipeline which brings us one step closer to a unified clustering-based method that can be trained end-to-end towards a diarization-specific objective.
\end{abstract}


\section{Introduction}
\vspace{-0.1cm}
Until a few years ago, competitive speaker diarization systems were mostly modular~\cite{sell2018diarization,landini2020but,park2019auto}, i.e., consisting of different modules to handle voice/speech activity detection (VAD/SAD), embedding extraction over uniform segmentation, clustering, optional resegmentation, and overlapped speech detection (OSD) and handling. However, end-to-end models such as end-to-end neural diarization (EEND)~\cite{fujita19_interspeech,horiguchi20_interspeech}, and two-stage systems such as target-speaker voice activity detection (TS-VAD)~\cite{medennikov2020target} or end-to-end with vector clustering (E2E-VC)~\cite{kinoshita2021integrating, kinoshita21_interspeech, Bredin23, Plaquet23} have recently gained more and more attention. The reasons for this are mainly their inherent ability to handle overlapped speech (where modular systems underperform) and fewer steps at inference time. Nevertheless, in contrast with modular systems, single-stage end-to-end systems do not handle well scenarios with many speakers~\cite{FITPT1357} and they are very data-hungry, requiring high volumes of training data with diarization annotations.

While two-stage systems produce per-frame speaker labels directly with a neural network (NN), they still build on clustering of embeddings: TS-VAD normally uses a clustering-based approach for initialization and recent competitive approaches based on E2E-VC~\cite{baroudi2023pyannote} make use of the best speaker embedding extractors available together with clustering to reconcile short-segment decisions. Besides, modular systems can still attain competitive performance in certain scenarios~\cite{landini2024diaper}, so speaker embedding extraction and clustering are still very relevant for diarization even today.

Speaker embedding extraction and clustering have been the main components of modular systems for more than a decade. Since the development of x-vectors~\cite{snyder2018x}, the embeddings have been NN-based with new versions such as ResNet~\cite{he2016deep,zeinali2019but}, ECAPA-TDNN~\cite{desplanques20_interspeech} or ECAPA2~\cite{thienpondt2023ecapa2} providing better and better results for speaker recognition and verification. These models are trained and utilized on at least a few seconds-long recordings for these tasks, but for diarization, embeddings are extracted on shorter segments since speakers can have short turns. However, the models are not designed for such usage, and tailoring them could lead to better performance~\cite{park2021multi,kwon2022multi,kim2023advancing,jung2023search,choi24d_interspeech}.

In the context of clustering-based diarization, VAD (and optionally OSD) is needed. 
In this work, we modify the embedding extractor to produce per-frame embeddings for the whole recording at once, naturally avoiding multiple calls to the embedding extraction routine and speeding up the process, while producing VAD and OSD labels for each embedding. This is done by removing the pooling mechanism and introducing linear layers to produce VAD and OSD decisions from the embeddings.
Moreover, we train the model for VAD+OSD and for embedding extraction on different data, thus taking advantage of different types of supervision that different corpora might offer, without generating synthetic training data, unlike end-to-end systems~\cite{fujita19_interspeech,landini22_interspeech,yamashita22_odyssey,landini2023multi}.

In related works like~\cite{miasato2017multi}, before x-vectors became popular, a single NN was used to extract per-frame speaker embeddings and produce VAD and OSD labels. 
However, the quality of the embeddings was restricted by the contrastive loss used to train the model and the limited speaker set contained in diarization-annotated datasets. In~\cite{cord2023frame}, per-frame embeddings were produced in a teacher-student framework where the teacher model produced per-segment embeddings.
More recently, in~\cite{kwon2021look,thienpondt24_odyssey}, the embedding extractor was used to provide VAD labels as a by-product by utilizing information encoded by intermediate representations in a weakly supervised VAD framework.

The results obtained with our proposed approach show that it is possible to train a single model for the three tasks (VAD, embedding extraction, OSD) and produce high-quality embeddings at a higher frequency. 
This opens up the space for building speaker verification systems that can discard silence- and overlap-related frames before producing per-utterance embeddings.
Moreover, the results are encouraging in our plan to combine this model with discriminative VBx (DVBx)~\cite{klement2024discriminative}, which will enable training of the whole modular pipeline in an end-to-end fashion towards a diarization-specific objective.

\section{Standard Methods}

\subsection{Diarization system pipeline}
\begin{figure}
\tikzstyle{block} = [rectangle, minimum height=0.8cm, minimum width=0.5cm, align=center,
      draw=none, shade,
      top color=black!20!blue!30,
      bottom color=blue!5,
      rounded corners=6pt,
      blur shadow={shadow blur steps=5}] 
\tikzstyle{arrow} = [single arrow, draw]
\tikzstyle{inv} = [rectangle, minimum height=0.1cm, minimum width=0.1cm] 
\tikzstyle{inv2} = [rectangle, minimum height=1cm, minimum width=0.1cm] 

\centering
\begin{subfigure}[b]{\linewidth}
\hspace{0.7cm}\resizebox{\columnwidth}{!}{%
    \begin{tikzpicture}[auto, >=stealth, node distance=0.4cm and 0.4cm, arr/.style={->,thick}, line/.style={thick}, font=\small]
    
    \node (VAD) [block, text width=0.6cm] {VAD};
    \node (emb) [block,text width=1.5cm, right=0.3cm of VAD] {Embedding extraction};
    \node (cluster) [block, text width=1.3cm, right=0.3cm of emb] {Clustering};
    \node (OSD) [block,text width=0.6cm , right=0.3cm of cluster] {OSD};
    \node (OVH) [block,text width=1.2cm , right=0.3cm of OSD] {Overlap handling};
    
    \node (pfeil0) [inv, left=0.3cm of VAD] {};
    \draw[arr] (pfeil0.east) -- (VAD.west) ;
    \draw[arr] (VAD.east) -- (emb.west);
    \draw[arr] (emb.east) -- (cluster.west);
    \draw[arr] (cluster.east) -- (OSD.west);
    \draw[arr] (OSD.east) -- (OVH.west);
    \draw[arr] (OVH.east) -- ++ (0.3,0.0) node[auto, xshift = 27]{};
    \end{tikzpicture}
}
   \caption{Standard pipeline.}
   \label{fig:standard_pipeline} 
\end{subfigure}

\centering
\begin{subfigure}[b]{\linewidth}
\hspace{0.9cm}\resizebox{0.95\columnwidth}{!}{%
    \begin{tikzpicture}[auto, >=stealth, node distance=0.4cm and 0.4cm, arr/.style={->,thick}, line/.style={thick}, font=\small]
    
    \node (emb) [block,text width=3.2cm] {Embedding extraction + VAD + OSD};
    \node (cluster) [block, text width=1.3cm, right=0.3cm of emb] {Clustering};
    \node (OVH) [block,text width=1.2cm , right=0.3cm of cluster] {Overlap handling};
    
    \node (pfeil0) [inv, left=0.3cm of emb] {};
    \draw[arr] (pfeil0.east) -- (emb.west) ;
    \draw[arr] (emb.east) -- (cluster.west);
    \draw[arr] (cluster.east) -- (OVH.west);
    \draw[arr] (OVH.east) -- ++ (0.3,0.0) node[auto, xshift = 27]{};
    \end{tikzpicture}
}
    \caption{Proposed pipeline.}
    \label{fig:proposed_pipeline}
\end{subfigure}

\caption{Comparison of modular diarization pipelines.}
\label{fig:pipelines}
\end{figure}
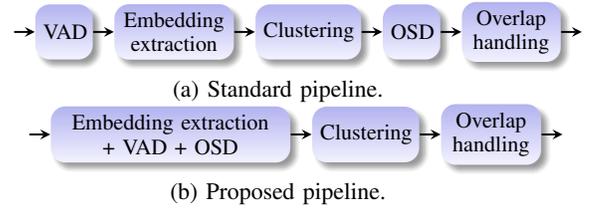

\vspace{-0.1cm}
We follow a standard modular pipeline (Figure~\ref{fig:standard_pipeline}), which requires VAD, embedding extraction, and clustering of those embeddings. Since clustering-based approaches assume a single speaker for each embedding, in order to handle overlaps between speakers, OSD is necessary. Given the overlap labels, we assign second speakers in those segments based on the heuristic~\cite{otterson2007efficient} that assigns the second closest (in terms of time) speaker. The embeddings and VAD and OSD labels can be produced by specific models trained for each of the tasks or, as we present in this work, produced by the joint speaker embedding extractor, VAD, and OSD model. The proposed pipeline is shown in Figure~\ref{fig:proposed_pipeline}. In both cases, we use VBx~\cite{landini2022bayesian} for clustering the embeddings with DVBx~\cite{klement2024discriminative} to find suitable hyperparameters for a given dataset.

\subsection{Baseline embedding extraction}
\begin{figure}
\centering
\begin{subfigure}{.33\linewidth}
  \centering
  \includegraphics[width=\linewidth]{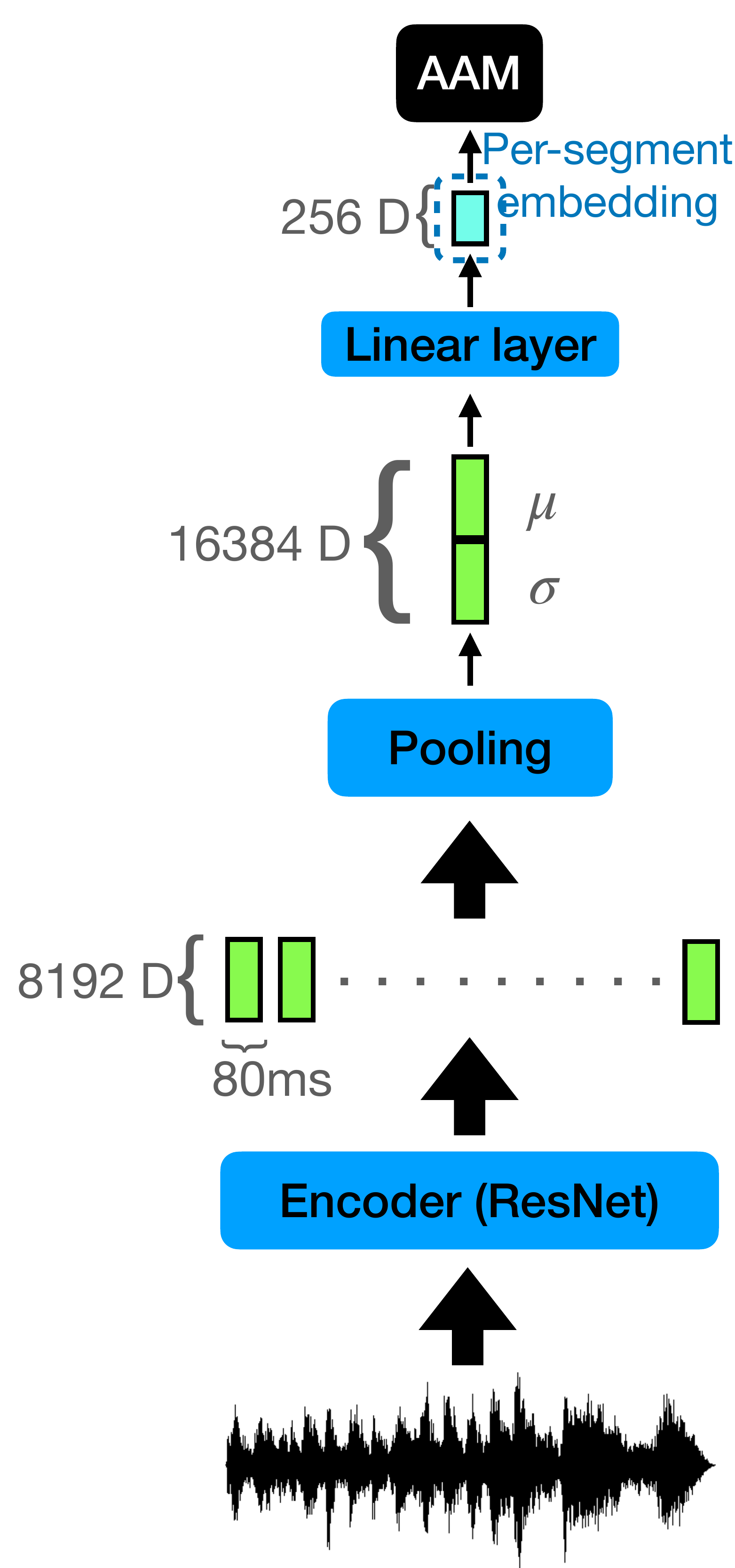}
  \caption{Per-segment.}
  \label{fig:extractor_original}
\end{subfigure}%
\begin{subfigure}{.66\linewidth}
  \centering
  \includegraphics[width=\linewidth]{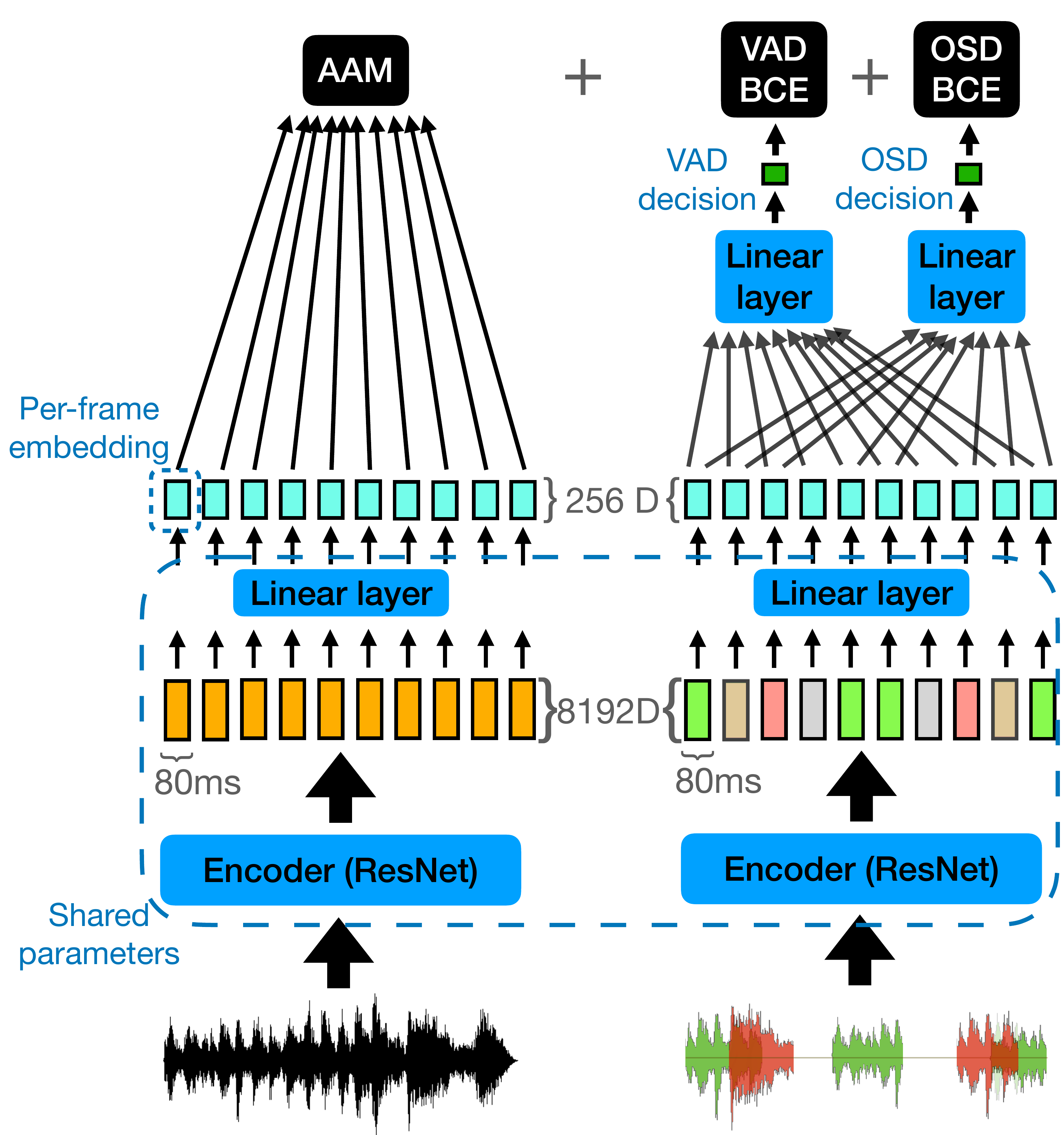}
  \caption{Per-frame (proposed).}
  \label{fig:extractor_proposed}
\end{subfigure}
\caption{Standard and proposed embedding extraction. In the proposed approach, all embeddings are used to calculate the VAD loss, but only those corresponding to speech are used for the OSD loss. Speaker embeddings are denoted in cyan.}
\label{fig:methods}
\end{figure}

\vspace{-0.1cm}
A typical embedding extractor used in speaker recognition (Figure~\ref{fig:extractor_original}) consists of an encoder processing the information frame-by-frame, a pooling mechanism, and a feed-forward NN processing segment-level information. In all our experiments, we utilize a ResNet-101 architecture~\cite{he2016deep} as an encoder for the embedding extractor~\cite{zeinali2019but}. However, the same ideas and extensions can be applied to other common architectures. The encoder transforms a sequence of 64-dimensional Mel-filterbanks extracted every 10\,ms into a shorter sequence (one vector per 80\,ms of the original audio) of internal 8192-dimensional representations. Note that the theoretical receptive field of ResNet101 is slightly longer than 2.5\,s
, so each of the internal representations is estimated on a relatively long segment of speech, much more than only 80\,ms. 

The encoder is followed by a pooling layer that combines the information along the temporal dimension. Thus, the whole audio is represented by a single fixed-dimensional vector independently of its length, thus ``per-segment'' embedding. As pooling layer, we chose the commonly used temporal statistical pooling~\cite{snyder2018x} - the concatenation of the mean and standard deviation of temporal representations. Then, the pooled vector is passed through a fully connected layer, which reduces its dimensionality, and finally enters the classification head during training. At inference time, the activations of the fully connected layer are used as speaker embeddings.

For the sake of making fair comparisons, all systems utilize a ResNet-101 trained with the WeSpeaker toolkit~\cite{WANG2024103104}. The models are trained to classify training speakers with AAM loss~\cite{deng2019arcface,xiang2019margin} on VoxCeleb2 (VC2)~\cite{chung18b_interspeech}. Training hyperparameters such as learning rate, margin scheduler, number of epochs, etc., follow the official WeSpeaker VoxCeleb recipe.

\section{Proposed Modifications}
\vspace{-0.1cm}
\subsection{High-resolution embedding extraction}
\vspace{-0.1cm}
The first modification over the usual architecture is the removal of the pooling layer common to all embedding extractors. We add a linear layer to reduce the dimensionality of the embeddings coming from the ResNet from 8192 to 256 - the same as our baseline embeddings. When doing so, the model simply transforms each of the frames into a speaker embedding, expected to have information about the speaker who spoke at that point in the audio. These low-dimensional representations are then used as input to the embedding extractor, as shown in Figure~\ref{fig:extractor_proposed} (left). The model is then trained with the same strategy as the original ``per-segment'' embedding extractor. The difference is that when training a ``per-frame'' embedding extractor, there is not a single embedding for the whole input audio but one embedding every 80\,ms - determined by the ResNet architecture.

While the parameters of the model can be learned from a random initialization, one of the alternatives that we explored for training the per-frame embedding extractor was to initialize the parameters of the encoder with those of a model trained to produce per-segment embeddings (removing the pooling layer), and then retrain it to produce per-frame embeddings. The obtained model produced embeddings that allowed for similar diarization performance as training the model without a pooling layer from scratch; however, the convergence was much faster. Note that this is different from the teacher-student approach followed in~\cite{cord2023frame}, where a student model is trained to produce higher resolution embeddings (yet still using pooling) given an already per-segment embedding extractor. In our case, we simply modify the original model and adapt it to produce higher-resolution embeddings.

\vspace{-0.2cm}

\subsection{Integrated VAD and OSD}
\vspace{-0.1cm}
The second modification, seen in Figure~\ref{fig:extractor_proposed} (right), is the addition of VAD and OSD ``heads'', which produce the probabilities that a given frame is speech and overlapped speech, respectively. Each per-frame embedding is passed through linear layers to produce the VAD and OSD logits, respectively. Both heads are trained using binary cross-entropy (BCE) loss. While all embeddings are used for the ``VAD loss'', only speech frames are used to calculate the ``OSD loss'', i.e., the head calculates the conditional probability that the frame is overlap given that it is speech.

In order to train the model in a multi-task fashion, the final loss to optimize is obtained as the weighted sum of the AAM, VAD BCE, and OSD BCE losses 
(with empirically obtained weights of 1, 5, and 2, respectively). Different aspects of the multi-task training need to be addressed. One of them is that, since VAD and OSD are, a priori, simpler tasks than speaker classification, there is no need to train the model on all tasks from the beginning. Furthermore, the same parameters are used for all tasks, but the model should use most of its potential for embedding extraction rather than VAD/OSD. For this reason, we first train the ResNet for per-frame embedding extraction. Only then the VAD and OSD heads are added and trained in a second training step for a few epochs. Training only for VAD and OSD (i.e., without the AAM loss) in the second stage can lead to worse speaker classification performance, so having the three losses is necessary. 

Another aspect is that the data used to train for embedding extraction usually consist only of the speech of a single speaker (i.e., no silence or overlaps). Hence, using the same data for training the VAD and OSD heads is not possible. For this reason, we utilize a compound set of different corpora with diarization annotations to train the VAD and OSD losses. Since these datasets usually do not have absolute speaker labels and also contain only a limited number of speakers, they are not suitable for optimizing the embedding extraction loss. Conversely, speaker labels are not necessary to compute the VAD and OSD losses. Therefore, in the multi-task training configuration, VC2 is used to calculate the AAM loss (Figure~\ref{fig:extractor_proposed} left), and the compound set is used to calculate the VAD and OSD losses (Figure~\ref{fig:extractor_proposed} right). 
At inference time, there is a single forward pass that produces embeddings and VAD and OSD decisions.
Note that this setup takes advantage of the supervision available for different tasks in a natural way and differs from EEND-like models which require all labels.

\section{Experimental Setup}
\vspace{-0.2cm}
\subsection{Datasets}
\vspace{-0.1cm}
In order to evaluate the robustness of the proposed approach, we utilized two popular datasets: DIHARD II~\cite{ryant2019second} and AMI~\cite{carletta2005ami} in its single distant microphone (SDM) version. 
Information about the datasets is presented in Table~\ref{tab:datasets_stats}.

To train the model for VAD and OSD, a compound set was utilized, comprised of data from AISHELL-4~\cite{fu21b_interspeech}, AliMeeting~\cite{yu2022m2met} (mix far and near field), AMI~\cite{carletta2005ami,kraaij2005ami} (mix-headset and mix-array), DIHARD III~\cite{ryant21_interspeech}, MSDWild~\cite{liu22t_interspeech}, and RAMC~\cite{yang22h_interspeech}. For DIHARD III, where a train set was not available, the development set was included in the compound set; otherwise, the train set was used. To train the model regarding the AAM loss, VoxCeleb2~\cite{chung18b_interspeech} was utilized, containing 2290 hours of speech from 5994 speakers.

\subsection{Baseline and proposed method configurations}
\vspace{-0.1cm}
The baseline system consists of VAD given by pyannote~\cite{Bredin23} followed by extraction of per-segment embeddings (1.5\,s long segments, every 0.25\,s) and clustering by means of VBx following the recipe in \cite{landini2022bayesian}. The embedding extractor and probabilistic linear discriminant analysis (PLDA) model needed by VBx were also trained on VC2 data. DVBx~\cite{klement2024discriminative} was used to obtain VBx hyperparameters tuned on the development (or train, in the case of AMI) set. As a final step, second speakers were assigned using OSD from pyannote and the heuristic that picks the closest in time speaker~\cite{otterson2007efficient}.

For the per-frame embedding-based systems, the configuration was the same as for the baseline, i.e., VBx hyperparameters were tuned on the dev set using DVBx. For the multi-task trained models, the VAD and OSD were obtained together with the per-frame embeddings. In all cases, the PLDA used by VBx was trained on VC2 using one random embedding per speaker selected from those produced by the corresponding model from randomly selected 6\,s of audio per utterance.

\section{Results}
\vspace{-0.1cm}
Systems are evaluated in terms of diarization error rate (DER): the sum of missed speech (Miss), false alarm speech (FA), and confusion (Conf.). VAD and OSD are evaluated in terms of misses and false alarms. All numbers are percentages. No forgiveness collar is used in any case.

\vspace{-0.2cm}

\subsection{Per-frame embeddings}
\vspace{-0.1cm}
The first modification introduced is regarding the change from per-segment to per-frame embeddings. The comparisons are presented in Table~\ref{tab:comparison_performance_VAD} where system (1) is the per-segment baseline. System (2) is trained to produce per-frame embeddings starting from randomly initialized parameters, while system (3) starts from the parameters of (1), removing the pooling layer and randomly initializing the last linear layer before producing the embeddings. To evaluate all three approaches, pyannote VAD is utilized and it is possible to see that they perform very similarly. It should be pointed out that 9.7\% out of the miss errors correspond to overlapped speech, which is not handled by these systems. While we expected the per-frame embeddings would lead to better results than the per-segment ones, the results did not necessarily prove this hypothesis. Per-frame embeddings represent only three times more frequent representations than the usual per-segment approach (80\,ms vs. 250\,ms) and it might be possible that such an increase is not enough to provide substantial gains. The effective receptive field might also impact the performance since non-speech parts of the audio might affect the embeddings and create a mismatch with the PLDA, which was trained only on speech. This should definitely be explored in the future, especially in connection with other architectures. 

Nevertheless, it should be mentioned that the proposed approach extracts all embeddings in a single call, unlike the standard per-segment approach where speech segments are fed and embeddings are extracted on short segments independently, resulting in several calls to the inference procedure. For a file 49.5\,min long, the per-segment method takes 28.7\,min to extract embeddings on one CPU, while the per-frame approach takes 9\,min (while jointly producing VAD and OSD labels).

\vspace{-0.3cm}

\subsection{Joint training}
\label{sec:join_experiments}
\vspace{-0.1cm}

System (5) in Table~\ref{tab:comparison_performance_VAD} continues training the embedding extractor from (3) but now using VC2 for the embedding extraction loss and the compound set for VAD and OSD losses. This model performs similarly to the ones previously described. The worse DER is mainly due to slightly worse VAD in comparison with pyannote's VAD possibly because they are trained on different compound sets. The intention here is not to directly contrast with pyannote but rather have a reference for comparison. However, these results show that it is possible to obtain competitive performance with joint training. In system (4), we explored using simulated conversations (SC)~\cite{landini22_interspeech} created from VC2 recordings to create training data with speaker, VAD and OSD labels. The VAD performance was extremely poor due to the large mismatch between SC and data manually annotated, thus a fine-tuning step on real data was necessary. For this reason, we utilized a compound set of real datasets for VAD and OSD. If pyannote VAD is used with the same embeddings, the DER can be lowered from 39.5 to 26.5, corresponding to 6.9 confusion error, showing that the quality of the embeddings is on par with (5).

Table~\ref{tab:comparison_performance_OSD} presents results after applying the overlap handling. System (7) corresponds to (1) after applying the heuristic with pyannote's OSD decisions and (8) corresponds to (5) after applying the heuristic with the OSD decisions from the proposed model. In both cases, there is a mild improvement in the overall DER of the same order for both systems. 

Finally, Table~\ref{tab:all_performance} compares (7) and (8) on DIHARD II and AMI (SDM). We can see that the proposed approach reaches a similar performance as the baseline. In the case of AMI, the VAD performance is better than what was observed for DIHARD II\footnote{AMI baseline: 10.2 miss, 1.9 FA and AMI proposed: 6.7 miss, 2.1 FA.}, most likely due to the inclusion of the train part of this dataset in the compound set (both for baseline and proposed). This is confirmed with (6) and (9), where (5) and (8), respectively, are fine-tuned on DIHARD dev (VAD and OSD tasks) and that improves the VAD performance.

\begin{table}
    \caption{Statistics of evaluation and compound training sets.}
    \label{tab:datasets_stats}
    \vspace{-2mm}
    \setlength{\tabcolsep}{5pt} % Default value: 6pt
    \centering
    \begin{tabular}{l|ccc|r}
    \toprule
    Dataset & Silence (\%) & 1-speaker (\%) & Overlap (\%) & Hours \\
    \midrule
   AMI (test) & 14.7 & 67.9 & 17.4 & 9.1 \\
   DIHARD2 (eval) & 25.9 & 67.5 & 6.6 & 22.5 \\
   \midrule
   Compound (train) & 24.3 & 55.0 & 20.7 & 774.2 \\
    \bottomrule
  \end{tabular}
\end{table}

\begin{table}[!tb]
  \centering
  \caption{Results on DIHARD II eval set. `S' means per-segment embeddings, and `F' means per-frame embeddings. `Comp.' means compound set. `FT' stands for finetuning.}
  \label{tab:comparison_performance_VAD}
  \vspace{-2mm}
  \setlength{\tabcolsep}{3pt} % Default value: 6pt
  \begin{tabular}{@{}
                  lll |
                  S[table-format=2.1]  
                  S[table-format=2.1]  
                  S[table-format=1.1] 
                  S[table-format=1.1] |
                  S[table-format=1.1] 
                  S[table-format=1.1]
                  @{}}
  \toprule
         \multicolumn{3}{c|}{System} & \multicolumn{4}{c|}{Diarization} & \multicolumn{2}{c}{VAD} \\
        
        \hspace{0.4cm}Type & Train data & VAD & 
        \multicolumn{1}{c}{DER} & 
        \multicolumn{1}{c}{Miss} & 
        \multicolumn{1}{c}{FA} &
        \multicolumn{1}{c|}{Conf.} &
        \multicolumn{1}{c}{Miss} &
        \multicolumn{1}{c}{FA} \\
  \midrule
  (1) S & VC2 & pyannote & 26.6 & 15.9 & 3.6 & 7.1 & 5.1 & 3.0 \\
  \midrule
  (2) F & VC2 & pyannote & 26.8 & 15.9 & 3.6 & 7.3 & 5.1 & 3.0 \\
  (3) S$\rightarrow$F & VC2 & pyannote & 26.4 & 15.9 & 3.6 & 6.9 & 5.1 & 3.0 \\
  \midrule
  (4) S$\rightarrow$F & SC VC2 & joint & 39.5 & 12.7 & 17.4 & 9.4 & 2.4 & 14.3 \\
  \midrule
  (5) S$\rightarrow$F & VC2. + Comp. & joint &  26.9 & 16.2 & 4.0 & 6.7 & 5.3 & 3.3 \\
  (6) S$\rightarrow$F & \hspace{0.05cm} FT VC2 + DH & joint & 26.1 & 15.0 & 3.5 & 7.5 & 4.4 & 2.9 \\
  \bottomrule
  \end{tabular}
\end{table}

\begin{table}[!tb]
  \centering
  \caption{Results after overlap post-processing on DIHARD II evaluation set. 
  `DH' stands for DIHARD II dev.}
  \label{tab:comparison_performance_OSD}
  \vspace{-2mm}
  \setlength{\tabcolsep}{3pt} % Default value: 6pt
  \begin{tabular}{@{}
                  lll |
                  S[table-format=2.1]  
                  S[table-format=2.1]  
                  S[table-format=1.1] 
                  S[table-format=1.1] |
                  S[table-format=1.1] 
                  S[table-format=1.1]
                  @{}}
  \toprule
         \multicolumn{3}{c|}{System} & \multicolumn{4}{c|}{Diarization} & \multicolumn{2}{c}{VAD} \\
        
        \hspace{0.4cm}Type & Train data & VAD & 
        \multicolumn{1}{c}{DER} & 
        \multicolumn{1}{c}{Miss} & 
        \multicolumn{1}{c}{FA} &
        \multicolumn{1}{c|}{Conf.} &
        \multicolumn{1}{c}{Miss} &
        \multicolumn{1}{c}{FA} \\
  \midrule
  (7) S & VC2 & pyannote & 26.2 & 14.2 & 4.5 & 7.5 & 5.2 & 0.7 \\
  \midrule
  (8) S$\rightarrow$F & VC2 + Comp. & joint & 26.6 & 15.3 & 4.5 & 6.8 & 5.9 & 0.4 \\
  (9) S$\rightarrow$F & \hspace{0.05cm} FT VC2. + DH & joint & 25.8 & 14.4 & 3.8 & 7.6 & 6.1 & 0.2 \\
  \bottomrule
  \end{tabular}
\end{table}

\begin{table}[!tb]
  \centering
  \caption{Performance comparison for per-segment embedding clustering using pyannote VAD and OSD versus per-frame embedding clustering using joint VAD and OSD.}
  \label{tab:all_performance}
  \vspace{-2mm}
  \setlength{\tabcolsep}{4.5pt} % Default value: 6pt
  \begin{tabular}{@{}
                  l |
                  S[table-format=2.1]  
                  S[table-format=2.1]  
                  S[table-format=2.1] 
                  S[table-format=2.1] | 
                  S[table-format=2.1]  
                  S[table-format=2.1]  
                  S[table-format=2.1]  
                  S[table-format=2.1] 
                  @{}}
  \toprule
        & \multicolumn{4}{c|}{Baseline} & \multicolumn{4}{c}{Proposed} \\
        
        Dataset       & 
        \multicolumn{1}{c}{DER} & 
        \multicolumn{1}{c}{Miss} & 
        \multicolumn{1}{c}{FA} &
        \multicolumn{1}{c|}{Conf.} &
        \multicolumn{1}{c}{DER} & 
        \multicolumn{1}{c}{Miss} & 
        \multicolumn{1}{c}{FA} &
        \multicolumn{1}{c}{Conf.} \\
  \midrule
  DIHARD II & 26.2 & 14.2 & 4.5 & 7.5 & 26.6 & 15.3 & 4.5 & 6.8 \\
  AMI (SDM) & 33.9 & 21.5 & 2.9 & 9.5 & 34.8 & 17.1 & 4.1 & 13.6 \\
  \bottomrule
  \end{tabular}
\end{table}

\section{Conclusions}
\vspace{-0.1cm}
In this work, we presented an approach to adapt a speaker embedding extractor for the purpose of diarization. Speaker embeddings were produced without a pooling mechanism while built-in mechanisms performed speech and overlap detection. With the proposed system, all three tasks are handled by a single model and the embedding extraction step is faster.

In spite of the competitive performance of the proposed approach, this work is an initial exploration  and we believe that many options are yet to be explored: 
\begin{itemize}[leftmargin=0.3cm]
    \item Use information from lower layers in the model since speech and overlap detection could better leverage it.
    \item ResNet is one of the most popular architectures for speaker embedding extractors, but other options need to be explored, such as variants based on time-delay NN. With the continuous development of foundation models, training a single model to perform different tasks at once should be possible. 
    \item Finally, this work should not be understood as a means in itself but rather as an intermediate goal towards other applications. On the one hand, for speaker verification, a system could automatically discard silence and overlap frames before producing (more robust) speaker embeddings without external VAD/OSD modules. On the other hand, we aim to combine the embedding extraction, speech, and overlap detection with DVBx in order to have a full pipeline that can be trained (or fine-tuned) in an end-to-end manner.
\end{itemize}

%\section*{Acknowledgment}

\bibliographystyle{IEEEtran}

\newpage
\bibliography{refs}

\begin{thebibliography}{10}
\providecommand{\url}[1]{#1}
\csname url@samestyle\endcsname
\providecommand{\newblock}{\relax}
\providecommand{\bibinfo}[2]{#2}
\providecommand{\BIBentrySTDinterwordspacing}{\spaceskip=0pt\relax}
\providecommand{\BIBentryALTinterwordstretchfactor}{4}
\providecommand{\BIBentryALTinterwordspacing}{\spaceskip=\fontdimen2\font plus
\BIBentryALTinterwordstretchfactor\fontdimen3\font minus
  \fontdimen4\font\relax}
\providecommand{\BIBforeignlanguage}[2]{{%
\expandafter\ifx\csname l@#1\endcsname\relax
\typeout{** WARNING: IEEEtran.bst: No hyphenation pattern has been}%
\typeout{** loaded for the language `#1'. Using the pattern for}%
\typeout{** the default language instead.}%
\else
\language=\csname l@#1\endcsname
\fi
#2}}
\providecommand{\BIBdecl}{\relax}
\BIBdecl

\bibitem{sell2018diarization}
G.~Sell \emph{et~al.}, ``{Diarization is Hard: Some Experiences and Lessons
  Learned for the JHU Team in the Inaugural DIHARD Challenge.}'' in
  \emph{Interspeech}, 2018, pp. 2808--2812.

\bibitem{landini2020but}
F.~Landini \emph{et~al.}, ``{BUT System for the Second DIHARD Speech
  Diarization Challenge},'' in \emph{International Conference on Acoustics,
  Speech and Signal Processing (ICASSP)}.\hskip 1em plus 0.5em minus
  0.4em\relax IEEE, 2020, pp. 6529--6533.

\bibitem{park2019auto}
T.~J. Park \emph{et~al.}, ``{Auto-tuning spectral clustering for speaker
  diarization using normalized maximum eigengap},'' \emph{IEEE Signal
  Processing Letters}, vol.~27, pp. 381--385, 2019.

\bibitem{fujita19_interspeech}
Y.~Fujita \emph{et~al.}, ``{End-to-End Neural Speaker Diarization with
  Permutation-Free Objectives},'' in \emph{Proc. Interspeech}, 2019.

\bibitem{horiguchi20_interspeech}
S.~Horiguchi \emph{et~al.}, ``{End-to-End Speaker Diarization for an Unknown
  Number of Speakers with Encoder-Decoder Based Attractors},'' in \emph{Proc.
  Interspeech}, 2020, pp. 269--273.

\bibitem{medennikov2020target}
I.~Medennikov \emph{et~al.}, ``{Target-Speaker Voice Activity Detection: A
  Novel Approach for Multi-Speaker Diarization in a Dinner Party Scenario},''
  in \emph{Proc. Interspeech}, 2020, pp. 274--278.

\bibitem{kinoshita2021integrating}
K.~Kinoshita, M.~Delcroix, and N.~Tawara, ``{Integrating end-to-end neural and
  clustering-based diarization: Getting the best of both worlds},'' in
  \emph{International Conference on Acoustics, Speech and Signal Processing
  (ICASSP)}.\hskip 1em plus 0.5em minus 0.4em\relax IEEE, 2021, pp. 7198--7202.

\bibitem{kinoshita21_interspeech}
------, ``{Advances in Integration of End-to-End Neural and Clustering-Based
  Diarization for Real Conversational Speech},'' in \emph{Proc. Interspeech},
  2021, pp. 3565--3569.

\bibitem{Bredin23}
H.~Bredin, ``{pyannote.audio 2.1 speaker diarization pipeline: principle,
  benchmark, and recipe},'' in \emph{Proc. INTERSPEECH 2023}, 2023.

\bibitem{Plaquet23}
A.~Plaquet and H.~Bredin, ``{Powerset multi-class cross entropy loss for neural
  speaker diarization},'' in \emph{Proc. INTERSPEECH 2023}, 2023.

\bibitem{FITPT1357}
\BIBentryALTinterwordspacing
F.~Landini, ``\BIBforeignlanguage{english}{{{From Modular to End-to-End Speaker
  Diarization}}},'' Ph.D. Thesis, Brno University of Technology, Faculty of
  Information Technology, 2024. [Online]. Available:
  \url{https://www.fit.vut.cz/study/phd-thesis/1357/}
\BIBentrySTDinterwordspacing

\bibitem{baroudi2023pyannote}
S.~Baroudi \emph{et~al.}, ``{pyannote. audio speaker diarization pipeline at
  VoxSRC 2023},'' \emph{The VoxCeleb Speaker Recognition Challenge 2023
  (VoxSRC-23)}, 2023.

\bibitem{landini2024diaper}
F.~Landini \emph{et~al.}, ``{DiaPer: End-to-End Neural Diarization with
  Perceiver-Based Attractors},'' \emph{IEEE/ACM Transactions on Audio, Speech,
  and Language Processing}, 2024.

\bibitem{snyder2018x}
D.~Snyder \emph{et~al.}, ``{X-vectors: Robust DNN embeddings for speaker
  recognition},'' in \emph{2018 IEEE international conference on acoustics,
  speech and signal processing (ICASSP)}.\hskip 1em plus 0.5em minus
  0.4em\relax IEEE, 2018, pp. 5329--5333.

\bibitem{he2016deep}
K.~He \emph{et~al.}, ``{Deep residual learning for image recognition},'' in
  \emph{Proceedings of the IEEE conference on computer vision and pattern
  recognition}, 2016, pp. 770--778.

\bibitem{zeinali2019but}
H.~Zeinali \emph{et~al.}, ``But system description to voxceleb speaker
  tecognition challenge 2019,'' \emph{arXiv preprint arXiv:1910.12592}, 2019.

\bibitem{desplanques20_interspeech}
B.~Desplanques, J.~Thienpondt, and K.~Demuynck, ``{ECAPA-TDNN: Emphasized
  Channel Attention, Propagation and Aggregation in TDNN Based Speaker
  Verification},'' in \emph{Proc. Interspeech 2020}, 2020.

\bibitem{thienpondt2023ecapa2}
J.~Thienpondt and K.~Demuynck, ``{ECAPA2: A hybrid neural network architecture
  and training strategy for robust speaker embeddings},'' in \emph{2023 IEEE
  Automatic Speech Recognition and Understanding Workshop (ASRU)}.\hskip 1em
  plus 0.5em minus 0.4em\relax IEEE, 2023.

\bibitem{park2021multi}
T.~J. Park, M.~Kumar, and S.~Narayanan, ``{Multi-scale speaker diarization with
  neural affinity score fusion},'' in \emph{ICASSP 2021-2021 IEEE International
  Conference on Acoustics, Speech and Signal Processing (ICASSP)}.\hskip 1em
  plus 0.5em minus 0.4em\relax IEEE, 2021, pp. 7173--7177.

\bibitem{kwon2022multi}
Y.~Kwon \emph{et~al.}, ``{Multi-scale speaker embedding-based graph attention
  networks for speaker diarisation},'' in \emph{ICASSP 2022-2022 IEEE
  International Conference on Acoustics, Speech and Signal Processing
  (ICASSP)}.\hskip 1em plus 0.5em minus 0.4em\relax IEEE, 2022, pp. 8367--8371.

\bibitem{kim2023advancing}
Y.~J. Kim \emph{et~al.}, ``{Advancing the dimensionality reduction of speaker
  embeddings for speaker diarisation: disentangling noise and informing speech
  activity},'' in \emph{ICASSP 2023-2023 IEEE International Conference on
  Acoustics, Speech and Signal Processing (ICASSP)}.\hskip 1em plus 0.5em minus
  0.4em\relax IEEE, 2023.

\bibitem{jung2023search}
J.-w. Jung \emph{et~al.}, ``{In search of strong embedding extractors for
  speaker diarisation},'' in \emph{ICASSP 2023-2023 IEEE International
  Conference on Acoustics, Speech and Signal Processing (ICASSP)}.\hskip 1em
  plus 0.5em minus 0.4em\relax IEEE, 2023, pp. 1--5.

\bibitem{choi24d_interspeech}
J.-H. Choi \emph{et~al.}, ``{Efficient Speaker Embedding Extraction Using a
  Twofold Sliding Window Algorithm for Speaker Diarization},'' in
  \emph{Interspeech 2024}, 2024, pp. 3749--3753.

\bibitem{landini22_interspeech}
F.~Landini \emph{et~al.}, ``{From Simulated Mixtures to Simulated Conversations
  as Training Data for End-to-End Neural Diarization},'' in \emph{Proc.
  Interspeech 2022}, 2022, pp. 5095--5099.

\bibitem{yamashita22_odyssey}
N.~Yamashita, S.~Horiguchi, and T.~Homma, ``{Improving the Naturalness of
  Simulated Conversations for End-to-End Neural Diarization},'' in \emph{Proc.
  The Speaker and Language Recognition Workshop (Odyssey 2022)}, 2022, pp.
  133--140.

\bibitem{landini2023multi}
F.~Landini \emph{et~al.}, ``{Multi-Speaker and Wide-Band Simulated
  Conversations as Training Data for End-to-End Neural Diarization},'' in
  \emph{ICASSP 2023-2023 IEEE International Conference on Acoustics, Speech and
  Signal Processing (ICASSP)}.\hskip 1em plus 0.5em minus 0.4em\relax IEEE,
  2023, pp. 1--5.

\bibitem{miasato2017multi}
V.~A. Miasato~Filho, D.~A. Silva, and L.~G.~D. Cuozzo, ``{Multi-objective
  Long-Short Term Memory Neural Networks for Speaker Diarization in Telephone
  Interactions},'' in \emph{2017 Brazilian Conference on Intelligent Systems
  (BRACIS)}.\hskip 1em plus 0.5em minus 0.4em\relax IEEE, 2017, pp. 181--185.

\bibitem{cord2023frame}
T.~Cord-Landwehr \emph{et~al.}, ``{Frame-wise and overlap-robust speaker
  embeddings for meeting diarization},'' in \emph{ICASSP 2023-2023 IEEE
  International Conference on Acoustics, Speech and Signal Processing
  (ICASSP)}.\hskip 1em plus 0.5em minus 0.4em\relax IEEE, 2023, pp. 1--5.

\bibitem{kwon2021look}
Y.~Kwon \emph{et~al.}, ``Look who’s not talking,'' in \emph{2021 IEEE Spoken
  Language Technology Workshop (SLT)}.\hskip 1em plus 0.5em minus 0.4em\relax
  IEEE, 2021, pp. 567--573.

\bibitem{thienpondt24_odyssey}
J.~Thienpondt and K.~Demuynck, ``{Speaker Embeddings With Weakly Supervised
  Voice Activity Detection For Efficient Speaker Diarization},'' in \emph{Proc.
  The Speaker and Language Recognition Workshop (Odyssey 2024)}, 2024, pp.
  131--136.

\bibitem{klement2024discriminative}
D.~Klement \emph{et~al.}, ``{Discriminative Training of VBx Diarization},'' in
  \emph{ICASSP 2024-2024 IEEE International Conference on Acoustics, Speech and
  Signal Processing (ICASSP)}.\hskip 1em plus 0.5em minus 0.4em\relax IEEE,
  2024, pp. 11\,871--11\,875.

\bibitem{otterson2007efficient}
S.~Otterson and M.~Ostendorf, ``Efficient use of overlap information in speaker
  diarization,'' in \emph{2007 IEEE Workshop on Automatic Speech Recognition \&
  Understanding (ASRU)}.\hskip 1em plus 0.5em minus 0.4em\relax IEEE, 2007, pp.
  683--686.

\bibitem{landini2022bayesian}
F.~Landini \emph{et~al.}, ``{Bayesian HMM Clustering of x-vector Sequences
  (VBx) in Speaker Diarization: Theory, Implementation and Analysis on Standard
  Tasks},'' \emph{Computer Speech \& Language}, vol.~71, p. 101254, 2022.

\bibitem{WANG2024103104}
\BIBentryALTinterwordspacing
S.~Wang \emph{et~al.}, ``Advancing speaker embedding learning: Wespeaker
  toolkit for research and production,'' \emph{Speech Communication}, vol. 162,
  p. 103104, 2024. [Online]. Available:
  \url{https://www.sciencedirect.com/science/article/pii/S0167639324000761}
\BIBentrySTDinterwordspacing

\bibitem{deng2019arcface}
J.~Deng \emph{et~al.}, ``{ArcFace: Additive Angular Margin Loss for Deep Face
  Recognition},'' in \emph{{Proceedings of the IEEE/CVF Conference on Computer
  Vision and Pattern Recognition (CVPR)}}, June 2019.

\bibitem{xiang2019margin}
X.~Xiang \emph{et~al.}, ``Margin matters: Towards more discriminative deep
  neural network embeddings for speaker recognition,'' in \emph{2019
  Asia-Pacific Signal and Information Processing Association Annual Summit and
  Conference (APSIPA ASC)}.\hskip 1em plus 0.5em minus 0.4em\relax IEEE, 2019,
  pp. 1652--1656.

\bibitem{chung18b_interspeech}
J.~S. Chung, A.~Nagrani, and A.~Zisserman, ``{VoxCeleb2: Deep Speaker
  Recognition},'' in \emph{Proc. Interspeech 2018}, 2018, pp. 1086--1090.

\bibitem{ryant2019second}
N.~Ryant \emph{et~al.}, ``{Second DIHARD challenge evaluation plan},''
  \emph{Linguistic Data Consortium, Tech. Rep}, 2019.

\bibitem{carletta2005ami}
J.~Carletta \emph{et~al.}, ``{The AMI meeting corpus: A pre-announcement},'' in
  \emph{International workshop on machine learning for multimodal
  interaction}.\hskip 1em plus 0.5em minus 0.4em\relax Springer, 2006, pp.
  28--39.

\bibitem{fu21b_interspeech}
Y.~Fu \emph{et~al.}, ``{AISHELL-4: An Open Source Dataset for Speech
  Enhancement, Separation, Recognition and Speaker Diarization in Conference
  Scenario},'' in \emph{Proc. Interspeech 2021}, 2021, pp. 3665--3669.

\bibitem{yu2022m2met}
F.~Yu \emph{et~al.}, ``{M2MeT: The ICASSP 2022 multi-channel multi-party
  meeting transcription challenge},'' in \emph{ICASSP 2022-2022 IEEE
  International Conference on Acoustics, Speech and Signal Processing
  (ICASSP)}.\hskip 1em plus 0.5em minus 0.4em\relax IEEE, 2022, pp. 6167--6171.

\bibitem{kraaij2005ami}
W.~Kraaij \emph{et~al.}, ``{The AMI meeting corpus},'' in \emph{Proc.
  International Conference on Methods and Techniques in Behavioral Research},
  2005.

\bibitem{ryant21_interspeech}
N.~Ryant \emph{et~al.}, ``{The Third DIHARD Diarization Challenge},'' in
  \emph{Proc. Interspeech 2021}, 2021, pp. 3570--3574.

\bibitem{liu22t_interspeech}
T.~Liu \emph{et~al.}, ``{MSDWild: Multi-modal Speaker Diarization Dataset in
  the Wild},'' in \emph{Proc. Interspeech 2022}, 2022, pp. 1476--1480.

\bibitem{yang22h_interspeech}
Z.~Yang \emph{et~al.}, ``{Open Source MagicData-RAMC: A Rich Annotated Mandarin
  Conversational(RAMC) Speech Dataset},'' in \emph{Proc. Interspeech 2022},
  2022, pp. 1736--1740.

\end{thebibliography}

\end{document}